\documentstyle[11pt,epsfig]{article}
\begin{document}

\title{RADII AND BINDING ENERGIES OF NUCLEI IN THE ALPHA-CLUSTER MODEL}

\author{G. K. NIE
\\
{\it Institute of Nuclear Physics, Ulugbek, Tashkent 702132, Uzbekistan}\\
galani@Uzsci.net}

\date{}
\maketitle

\begin{abstract}

The $\alpha$-cluster model is based on two assumptions that the proton-neutron pair
interactions are responsible for the adherence between $\alpha$-clusters and that
NN-interactions in the $\alpha$-cluster structure are isospin independent. It allows
one to estimate the Coulomb energy and the short range inter-cluster bond energy in
dependence on the number of clusters. Charge radii are calculated on the number of
$\alpha$-clusters too. Unlike the Weizs\"{a}cker formula in this model the binding
energies of alpha-clusters and excess neutrons are estimated separately. The
calculated values are in a good agreement with the experimental data.

{{\it keywords}: nuclear structure; alpha-cluster model; Coulomb energy; Surface
tension energy, binding energy; charge radius}

\end{abstract}

{\it{PACS} Nos.: 21.60.-n; 21.60.Gx; 21.10.Dr; 21.60.Cs.}

\section{Introduction}

The liquid drop model is explicitly formulated by two phenomenological formulas, the
Weizs\"{a}cker formula [1] to calculate the nuclear binding energy and the formula
$constA^{1/3}$ to calculate the nuclear radii. It is known [2] that the idea of
alpha clustering has developed together with the liquid drop model since the
beginning of the nuclear physics. Nowadays the alpha-cluster model is becoming more
popular. Recently it has been shown that a large body of data on the $\alpha$-decay
is described by a simple cluster model with using a notion of an $\alpha$-core
potential [3]. There are several attempts to incorporate the alpha-cluster model
into the shell model [2,4]. To elucidate the optical potential in nucleus-nucleus
collisions an idea of bigger clusters like nucleon molecules of $^{12}\rm C$ has
been used [5]. Currently the idea of alpha-clustering is actively used in the
microscopic studies of the nuclear structure of the light nuclei in the framework of
the few-body approach [6], as well as in the microscopic approaches to the effective
nucleon-nucleon interaction [7]. Also there have been some studies recently to apply
an alpha-cluster model to facilitate the description of some characteristics of
nuclei like the $\alpha$-particle separation energy [8] and the nuclear radii [9]
for a wide range of nuclei.

There is another strong evidence in favor of an alpha-cluster model of nuclear
structure. It comes from the recent findings [10,11] that the radii of nuclei can be
described independently of the number of excess neutrons by the formula $R = const
(Z)^{1/3}$ fm. It can be considered as a consequence of a well known fact that the
charge density in nuclei is almost constant (see, for example, Table 3.7 [12]).

In this connection it is interesting to obtain the formulas to calculate the charge
radii and binding energies of nuclei on the basis of the idea that the nuclear
structure is defined by its elements, $\alpha$-clusters. In such an approach, unlike
the Weizs\"{a}cker formula, the formula to calculate the nuclear binding energy
should consist of two separate parts. One is responsible for the binding energy of
$\alpha$-clusters and the other should estimate the binding energy of excess
neutrons.

The core starts growing with $Z$ from the nucleus with $Z \geq 10$ [10] together
with the surface tension energy. The distance between the nearby clusters is
determined by their charge radii, which are bigger than their matter radii. It was
shown [10,11] that the $\beta$-stable nuclei have a particular number of excess
neutrons needed to fill in the difference in the volumes occupied by the charge and
the matter of the nucleons of the $\alpha$-clusters in the core. Then the radius
$R_m$ of a $\beta$-stable isotope can be estimated by the volume occupied by the
matter of the core and the volume of the charge of a few peripheral clusters. It has
been shown that the condition $R_m=R$ determines the narrow $\beta$-stability path
and its width. The binding energy of these excess neutrons can be calculated as a
sum of the energies of $nn$-pairs.

The hypothesis that the $nn$-pairs are placed in the core explains its high mass
density. In addition, it provides an explanation [10] why the experimental values of
the most abundant isotopes $R^{abn}_{exp}$ are well described by both functions,
$\sim A^{1/3}$ in the liquid drop model and $\sim Z^{1/3}$ in the $\alpha$-cluster
model. Then it is expected that the NN-interaction in the regular $\alpha$-cluster
structure is different from that in excess neutron matter.

It is approved by a recent investigation of the optical potential for the elastic
channels in reaction $^{48}\rm Ca (p,n)^{48}\rm Sc$ [13]. It has been found that the
projectile proton is involved in the NN-interactions in $p+^{48}\rm Ca$ channel
somewhat different from NN-interactions of the outgoing excess neutron in
$n+^{48}\rm Sc$ channel. The proton belongs to the $\alpha$-cluster structure and
the neutron is from the excess neutrons. To elucidate the isospin independence of
NN-interactions between the nucleons of the $\alpha$-cluster structure the same
investigation should be done for the different reaction $^{41}\rm Ca (p,n)^{41}\rm
Sc$ where the both proton and neutron belong to one $pn$-pair.

The energy of the short-range inter-cluster bonds  $\epsilon_{\alpha \alpha}$ is
taken to be equal $\epsilon_{\alpha \alpha} = \epsilon^{nuc}_{\alpha \alpha} -
\epsilon^{C}_{\alpha \alpha}$ where $\epsilon^{nuc}_{\alpha \alpha}$ denotes the
nuclear force energy and $\epsilon^{C}_{\alpha \alpha}$ denotes the Coulomb
repulsion energy between two nearby clusters. The old term 'nuclear force energy'
means here the energy of the attraction between nearby clusters due to the strong
inter-nucleon potential. The values of $\epsilon_{\alpha \alpha}=2.425$ MeV,
$\epsilon^{nuc}_{\alpha \alpha}=4.350$ MeV and $\epsilon^{C}_{\alpha \alpha} =1.925$
MeV have been found from analysis of the experimental data of the lightest nuclei
with $Z \leq 6$ [14] under two assumptions that the proton-neutron ($pn$) pair
interactions are responsible for the adherence between $\alpha$-clusters and that
the proton and the neutron belonging to one pair have equivalent single-particle
nuclear bound state potentials, the EPN requirement [14,15]. The latter allows one
to take the difference between the single-particle binding energies of the proton
and the neutron of the last pair $\Delta E_{pn}$ as the  Coulomb energy of the
proton's interaction with the other protons of the nucleus.

It was shown [14,15] that the experimental binding energies $E_{\exp}$ of the
$\alpha$-cluster nuclei with $N=Z \geq 6$, known for the nuclei with $Z \leq 28$,
are well described by the formula $E^b = N_{\alpha}\varepsilon_\alpha + 3(N_{\alpha}
- 2)\varepsilon_{\alpha \alpha}$ where $\varepsilon_\alpha=\epsilon_{4\rm
He}=28.296$ MeV. The number of bonds $3(N_\alpha-2)$ means that every new cluster
brings three bonds with the nearby clusters. This idea was discussed before [2].  In
the nuclei heavier than $^{16}\rm O$ the long-range part of the Coulomb interaction
of the last $\alpha$-cluster with the remote clusters starts growing. It must be
compensated by its surface tension energy $E^{st}_{\alpha}$. Then the empirical
values of the surface tension energy of the last $\alpha$-cluster $E^{st}_{\alpha}$
can be obtained from Eq. $E^{st}_{\alpha}= \Delta E_{\alpha} - (\epsilon^C_{\alpha}
+ 3\epsilon^{C}_{\alpha \alpha})$ where $\epsilon^C_{\alpha}$ stands for the
internal Coulomb energy of one $\alpha$-cluster, which equals the Coulomb energy of
the nucleus $^4\rm He$ $\epsilon^C_{\alpha}=0.764$ MeV, and  $\Delta
E_{\alpha}=\sum^2 \Delta E_{pn}$ of two $pn$-pairs belonging to one cluster
[14,16,17].

The long-range part of the Coulomb energy of the last cluster's interaction with
$N_{\alpha}-4$ remote clusters is well approximated by expression
$2(Z-8)e^2/(1.2R_p)$ where $R_p$ is the radius of the last proton position in the
nucleus. Then it is expected that $E^{st}_{\alpha} = 2(Z-8)e^2/(1.2R_p)=\gamma_1
R_p^2$, which determines $R_p$. To find the Coulomb radius $R_C$, the approximation
$E^{st}_{\alpha}=\gamma_1 R_C^2 =\gamma_2 N^{2/3}_{\alpha}$ is used. From the
analysis of the empirical values of $\Delta E_{\alpha}$ known for the nuclei with $Z
\leq 22$ with an accuracy of a few KeV the values of $\gamma_1$ and $\gamma_2$ have
been obtained together with the formulas to calculate $R_p$ and $R_C$ on the number
$N_\alpha$.

Thus, the formula to calculate the binding energy with using the nuclear force
energy of $\alpha$-clusters $E_{\alpha}^{nuc}=N_{\alpha}\epsilon^{nuc}_{\alpha}
+3(N_{\alpha}-2)\epsilon^{nuc}_{\alpha \alpha}$, the Coulomb energy $E^C = 3/5
(Ze)^2/R_C$ and the surface tension energy $E^{st}=\sum^{N_{\alpha}}_5
E^{st}_\alpha$ plus the binding energy of the excess neutrons $E_{\Delta N}$
estimated as a sum of separation energies of $nn$-pairs describes the binding energy
of the nuclei with $N_\alpha \geq 5$, with the errors close to those of the
Weizs\"{a}cker formula [16,17].

It is evident that the model can be applied only for the nuclei with the number of
excess neutrons $\Delta N \geq 0$, so that every proton is coupled with a neutron.
This set of nuclei includes all stable nuclei and the unstable nuclei around the
$\beta$-stability path. The model has been developed for the last few years and the
main formulas have been published [10,11,14,15,16,17]. In the most recent papers
[16,17] the surface tension energy was calculated with using the empirical values of
$E_\alpha^{st}$ for $Z \leq 28$, $N_\alpha \leq 14$. In this paper the value
$E^{st}$ is calculated  by one formula for $9 \leq Z \leq 118$. Besides, the
phenomenological formula to calculate the excess neutron binding energy has been
extended for the cases of odd number of excess neutrons.

\section{Nuclear Binding Energy }
\subsection{Nuclear force energy, Coulomb energy and surface tension energy}

The picture given in Fig. 1  shows the $pn$-pair bonds in the lightest nuclei.
\begin{figure}[th]
\centerline{\psfig{file=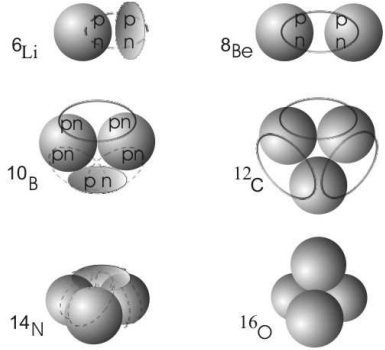,width=5cm}} \vspace*{8pt} \caption{A schematic
illustration of two kinds of bonds between nearby $pn$-pairs, which define
short-range inter-cluster bonds (solid line ellipse) and the bonds between the
single $pn$-pair in the nuclei with odd $Z$ and the $pn$-pairs of nearby
$\alpha$-clusters(dashed line ellipse).}
\end{figure}
The distance between clusters is supposed to be constant for all nuclei except the
nucleus $^8\rm Be$. In this case one bond is not enough to keep two positively
charged clusters close. Adding one more $pn$-pair causes two new bonds with the
pairs of two alpha-clusters with the energy $2\varepsilon_{pn pn}$ in the nucleus
$^{10}\rm B$, and the clusters get closer with the distance proper for the
alpha-cluster liquid. The binding energy of the lightest nuclei $E^b$ is calculated
in accordance with Fig. 1  by the following equations:
\begin{eqnarray}
\nonumber
  E^b(4\rm He) &=& \varepsilon_\alpha,\\
\nonumber
  E^b(6\rm Li) &=& \varepsilon_\alpha + \varepsilon_{pn pn} + \varepsilon_{pn}, \\
\nonumber
  E^b(10\rm B) &=& 2\varepsilon_\alpha + \varepsilon_{\alpha \alpha} + 2\varepsilon_{pn pn} + \varepsilon_{pn}, \\
\nonumber
  E^b(12\rm C) &=& 3\varepsilon_{\alpha}+3\varepsilon_{\alpha \alpha},
\end{eqnarray}
where $\varepsilon_{\alpha \alpha}$ denotes the binding energy between two nearby
alpha-clusters, $\varepsilon_{pn pn}$ denotes the binding energy between the single
$pn$-pair and a $pn$-pair of the nearby $\alpha$-clusters and $\varepsilon_{pn}$
stands for the binding energy of the proton and the neutron in the single pair.
Solving these four equations with the values of $E^b$ equal to the experimental
values $E_{exp}$ [18] with an accuracy of 1 KeV (at experimental measurement error
within 2-3 KeV) gives the values $\varepsilon_\alpha = 28.296$ MeV,
$\varepsilon_{\alpha \alpha} = 2.425$ MeV, $\varepsilon_{pn pn} = 2.037$ MeV,
$\varepsilon_{pn} =1.659$ MeV.

The experimental values of the other nuclei including $^{12}\rm C$ are well
approximated by means of the following equations for the nuclei with even $Z$ with
the mass number $A$:
\begin{equation}
E^b = N_{\alpha}\varepsilon_\alpha + 3(N_{\alpha} - 2)\varepsilon_{\alpha \alpha},
\label{1}
\end{equation}
and for the nuclei with odd $Z_1 = Z + 1$ with the mass number $A_1 = A + 2$

\begin{equation}
E^b_1 = E^b + 6\varepsilon_{pn pn} + \varepsilon_{pn}. \label{2}
\end{equation}

The energy of nucleus $A_1$ is determined by the energy of the nucleus $A$ plus the
energy of six bonds between the single $pn$-pair and the six $pn$-pairs of the three
nearby clusters plus the binding energy of the pair itself. The values of $E^b$ are
presented in Table 1 in comparison with the experimental values $E_{exp}$.

\begin{center}{Table 1. Binding energies calculated for the nuclei with $N=Z$.
The values are given in MeV. Empirical values are given to three places of decimals
and calculated ones are given to one place of decimals.}\end{center}
{\begin{tabular}{ccccccccccc}\hline $Z$&$E_{exp}$[18]&$E^b$(1,2)&$\Delta
E_{pn}$&$\Delta E_\alpha$&$E^C$(3,4)&$E_{\alpha}^{nuc}$(5)&$E^{st}$(6)&$\Delta
E^b$(7)&$\Delta E^{Wz}(10)$
\\\hline
 2& 28.296&28.296 & .764&  .764 &  .764 &  29.060 & 0.000&0.0 &-.1 \\
 3& 31.992&31.992 &1.007&       & 1.771 &  33.753& 0.000&0.0  &8.7 \\
 4& 56.498&56.498 &1.644& 2.651 & 3.415 &  59.912& 0.000&0.0  &0.2 \\
 5& 64.750&64.750 &1.850&       & 5.303 &  70.205& 0.000&0.0  &4.2 \\
 6& 92.163&92.163 &2.764& 4.614 & 8.067 & 100.230 & 0.000&0.0 &2.4 \\
 7&104.659&106.0&3.003&      & 11.070& 117.1   & 0.000&-1.3   &6.7  \\
 8&127.620&127.7&3.536& 6.539& 14.606& 142.3    & 0.000&-0.1  &2.4  \\
 9&137.371&141.6&3.544&      & 18.150& 159.2&0.541  &0.2 &2.0   \\
10&160.646&163.3&4.021& 7.565& 22.171& 184.5& 1.026& 2.2&-0.6  \\
11&174.146&177.2&4.328&      & 26.499& 201.3& 2.351& 2.0&1.4  \\
12&198.258&198.9&4.838& 9.166& 31.337& 226.6& 3.653& 1.9& 0.7  \\
13&211.896&212.8&5.042&      & 36.379& 243.4& 5.692& 1.8&2.1  \\
14&236.536&234.4&5.593&10.635& 41.972& 268.7& 7.749& 2.7&2.7   \\
15&250.603&248.3&5.731&      & 47.703& 285.6&10.477& 3.1&4.0   \\
16&271.780&270.0&6.227&12.958& 53.930& 310.8&13.169& 1.1&1.9   \\
17&285.573&283.9&6.351&      & 60.281& 327.7&16.517& 1.3&2.6   \\
18&306.719&305.6&6.746&13.097& 67.027& 352.9&19.727&-0.4&1.1   \\
19&320.634&319.5&6.923&      & 73.950& 369.8&23.647&  0.0&1.8   \\
20&342.056&341.2&7.286&14.209& 81.236& 395.0&27.397&-1.0&1.1   \\
21&354.710&355.0&7.277&      & 88.513& 411.9&31.671&-1.8& 0.4   \\
22&375.587&376.7&7.615&14.892&96.128 & 437.1&35.750&-2.9&-0.2   \\
23&390.350&390.6&*7.7&        &103.8 & 454.0&40.5  &-1.6&1.1   \\
24&411.720&412.3&8.165&15.9  &112.0  & 479.2&45.1  &-1.8&1.5   \\
25&426.659&426.2&8.496&      &120.5  & 496.1&50.6  & -0.2&3.0   \\
26&447.669&447.9&*8.4&16.9    &128.9 & 521.3&55.5  & -0.5&3.5   \\
27&462.726&461.8&*8.4&        &137.3 & 538.2&60.9  & 1.3&5.2   \\
28&484.010&483.4&*8.8&17.2    &146.1 & 563.4&66.1  & 1.6&6.4   \\
29&497.111&497.3&9.259&      &154.8  & 580.3&71.8  & 1.6&6.2   \\\hline
\end{tabular}}

The other values in Table 1 have been defined in Introduction and are provided by
the number of the corresponding formula. The values for even $Z$ and for odd $Z_1$
nuclei, namely $E^b$ and $E^b_1$, $E^C$ and $E^C_1$, $E_{\alpha}^{nuc}$ and
$E_{\alpha 1}^{nuc}$, $E^{st}$and $E^{st}_1$ are indicated as $E^b$,
$E_{\alpha}^{nuc}$, $E^{st}$ for the nuclei with $Z$ with the mass number $A$. The
9th column titled as '$\Delta E^b$(7)' gives the difference between $E_{exp}$ and
$E^b$ calculated by (7). The column with '$\Delta E^{Wz}$(10)' contains the
differences between $E_{exp}$ and $E^{Wz}$ calculated by the Weizs\"{a}cker
formula(10). The values indicated by '*' have been obtained by means of (17)/(18)
and (21), see Section 3,  with the accuracy of several tenth of MeV.

The EPN requirement comes from isospin independence of NN-interactions in
$\alpha$-clusters. It provides a possibility to estimate the total Coulomb energy of
a nucleus, $E^C = \sum \Delta E_{pn}$, where $\Delta E_{pn}$ is the differences
between the binding energies of the proton and neutron belonging to one $pn$-pair.
If a sum of the values $\Delta E_{pn}$ of two $pn$-pairs belonging to one
$\alpha$-cluster is presented as the Coulomb energy of the $\alpha$-cluster $\Delta
E_{\alpha} = \sum^2 \Delta E_{pn}$, then

\begin{equation}
E^C = \sum^{N_{\alpha}}_1\Delta E_{\alpha} + \delta,\label{3}
\end{equation}
where, the value of $\delta$ comes from taking into account that two clusters in
$^8\rm Be$ are at bigger distance than in $^{10}\rm B$, see Fig. 1. For the nucleus
$A_1$ the corresponding equation is

\begin{equation}
E^C_{1} = \sum^{N_{\alpha}}_1\Delta E_{\alpha} + \delta + \Delta E_{pn}. \label{4}
\end{equation}

The values of $\Delta E_{pn}$ for the nuclei with $N = Z$ are obtained by using the
experimental binding energies taken from [18,19]. The binding energies of the nuclei
with $(Z-1,N)$ and $(Z,N-1)$ needed for calculation of $\Delta E_{pn}$ for the
nuclei with $3 \leq Z \leq 8$ and $Z = 24, 25$ have been calculated from the mass
deficiency [19]. The energy of $^4\rm He$ 28.296 MeV and of $^5\rm He$ 27.338 MeV
are taken from [18]. For the nuclei with $Z = 23, 26, 27, 28$ the values of $\Delta
E_{pn}$ cannot be obtained by this procedure due to the lack of experimental data
for the nuclei with $(Z, N-1)$. The values have been estimated in the framework of
this model by equations (17)/(18) with using the values of the radius of the last
proton position $R_{p/p1}$ (21), see Section 3.

The value of the Coulomb repulsion energy between two nearby clusters
$\varepsilon^C_{\alpha \alpha} = 1.925$ MeV has been found from the equation for the
Coulomb energy of one cluster in $^{12}\rm C$ $\Delta E_\alpha(^{12}\rm C) =
\varepsilon ^C_\alpha + 2\varepsilon ^C_{\alpha \alpha}$, where
$\varepsilon^C_\alpha$ denotes the internal Coulomb energy of one cluster
$\varepsilon^ C_\alpha = E^C_{4\rm He} = 0.764$ MeV. This value perfectly  fits the
corresponding equation for $^{16}\rm O$, which is  $\Delta E_\alpha(^{16}\rm O) =
\varepsilon ^C_\alpha + 3\varepsilon ^C_{\alpha \alpha}$. The value of the Coulomb
repulsion energy between the single $pn$-pair and one nearby alpha-cluster in case
of odd $Z_1$ is found from the data for the nuclei $^6\rm Li$ and $^{14}\rm N$,
$\varepsilon ^C_{pn \alpha} = 1.001(6)$ MeV.

One can estimate the Coulomb repulsion energy between two alpha clusters in the
$^8\rm Be$ as follows, $\varepsilon ^C_{\alpha \alpha}(^8\rm Be) = \ E^C_{^8\rm Be}
- 2\varepsilon ^C_\alpha = 1.887$ MeV, which is less than $\varepsilon ^C_{\alpha
\alpha}$ by the value of $\delta = 0.038$ MeV.

Eqs. (3) and (4)  give the empirical values of the total Coulomb energy $E^C$
obtained on the basis of the values known from the experimental data with an
accuracy of a few KeV. Having obtained the total Coulomb energy of nuclei as a sum
of $\Delta E_{pn}$ of the pairs, one can easily test the validity of the hypothesis
of the alpha-cluster structure on the nuclei with few clusters $N_\alpha \leq 4$,
that is the nuclei where each cluster is in touch with everyone of the other
clusters. In this case the Coulomb energy can be calculated by the number of
clusters and their bonds.

For $^8\rm Be$ according to Fig. 1.  $E^C(^8\rm Be)= 2\varepsilon ^C_\alpha +
\varepsilon ^C_{\alpha \alpha}(^8\rm Be)$. For $^{12}\rm C$ and $^{16}\rm O$
$E^C(^{12}\rm C)= 3\varepsilon ^C_\alpha + 3\varepsilon ^C_{\alpha \alpha}$ and
$E^C(^{16}\rm O)= 4\varepsilon ^C_\alpha + 6\varepsilon ^C_{\alpha \alpha}$. In case
of $^6\rm Li$ $E^C(^6\rm Li) =\varepsilon ^C_\alpha + \varepsilon ^C_{pn \alpha}$
and for $^{10}\rm B$ $E^C(^{10}\rm B) = 2\varepsilon^C_\alpha + \varepsilon
^C_{\alpha \alpha} + \Delta E_{pn} $. For $^{14}\rm N$  $E^C(14{\rm N}) = E^C(12\rm
C) + 3\varepsilon ^C_{pn \alpha}$. For these  nuclei the equations give values of
$E^C$ in agreement with the values obtained by (3) and (4), see Table 1, with an
accuracy of 1 KeV, for  $^6\rm Li$ with an accuracy of 6 KeV. It confirms the
validity of the cluster structure of the nuclei with the quantities of $\varepsilon
^C_{\alpha \alpha}$ and $\varepsilon ^C_{pn \alpha}$, as well as the equations (3)
and (4) are confirmed as the reliable formulas for obtaining the empirical values of
the Coulomb energy for the nuclei with $Z \leq 22$ with the sure accuracy  of few
tens KeV. It also approves the statement about isospin independence of
NN-interaction.

After adding the values of $\Delta E_{pn}$ for the nuclei with $Z=26,28$ and $Z=29$
estimated in the framework of the model with an accuracy of several tenth MeV (the
accuracy is determined by the difference in the values obtained by the other
possible way, by using (19)/(20), see section 3), the accuracy of the value of $E^C$
for $^{29}\rm Cu$ is expected to be about 1 - 2 MeV.

One can estimate the nuclear force energy of $\alpha$-clusters from Eq.
$\varepsilon^{nuc}_\alpha = \varepsilon _\alpha + \varepsilon ^C_\alpha = 29.060$
MeV and $\varepsilon ^{nuc}_{\alpha \alpha} = \varepsilon _{\alpha \alpha} +
\varepsilon ^C_{\alpha \alpha} = 2.425 \rm MeV + 1.925 MeV = 4.350\rm MeV $. This
consideration allows one to estimate $\varepsilon^{nuc}_{\alpha \alpha}(^8\rm Be) =
(E_{exp}(^8\rm Be)-2\varepsilon_\alpha)+\varepsilon ^C_{\alpha \alpha}(^8\rm Be) =
1.792$ MeV.

The empirical values of nuclear force energy of $\alpha$-clusters $E_{\alpha}^{nuc}$
for the lightest nuclei $Z \leq 6$, $N_\alpha \leq 3$, are obtained from
$E_{exp}=E^{nuc}+E^C$ and they are presented in Table 1. For the nuclei with $Z \geq
7$, $N_\alpha \geq 3$, one gets

\begin{equation}
E_{\alpha}^{nuc} = N_\alpha \varepsilon ^{nuc}_\alpha + 3(N_\alpha
-2)\varepsilon^{nuc}_{\alpha \alpha}  ; E_{\alpha 1}^{nuc} = E^{nuc} +
6\varepsilon^{nuc}_{pn pn} + \varepsilon_{pn},\label{5}
\end{equation}
where $6\varepsilon^{nuc}_{pn pn} + \varepsilon_{pn} = 6\varepsilon_{pn pn} +
3\varepsilon^C_{pn \alpha} + \varepsilon_{pn} = 16.884$ MeV.

The Coulomb energy for the nuclei with $N_\alpha > 4$ consists not only of the
energies of bonds between nearby clusters. It must contain the long-range Coulomb
interaction part. That fact that Eq. (1)/(2) has a good agreement with the
experimental values allows one to suggest that the long-range part of the Coulomb
energy $E^{C_{lr}}_\alpha$ between the last $\alpha$-cluster and the $N_\alpha - 4$
remote clusters is compensated by the surface tension energy, which means that the
$E^{st}\neq 0$ for the nuclei with $Z \geq 9$.  It gives an opportunity to obtain
the empirical values $E^{st}$ from the values of $\Delta E_{\alpha}$ and $\Delta
E_{pn}$, known with a good accuracy.

The Coulomb energy of the last $\alpha$-cluster for the even nuclei with $N_\alpha
\geq 5$ is defined as $\Delta E_\alpha = \varepsilon^C_\alpha + 3\varepsilon
^C_{\alpha \alpha} + E^{Clr}_\alpha$ and it is suggested that the surface tension
energy of the last cluster $E^{st}_\alpha = E^{Clr}_\alpha$. The Coulomb energy of
the single $pn$-pair in case of odd $Z_1$ $\Delta E_{pn} = 3\varepsilon
^C_{pn\alpha} + E^{Clr}_{pn}$ and the surface tension energy of the $pn$-pair
$E^{st}_{pn} = E^{Clr}_{pn}$. Therefore, the empirical values of the surface tension
energy are obtained from

\begin{equation}
E^{st} = \sum_{5}^{N_\alpha}E^{st}_\alpha ; E^{st}_1 = E^{st} + E^{st}_{pn},
\label{6}
\end{equation}
where $E^{st}_\alpha= \Delta E_\alpha - \varepsilon^C_\alpha - 3\varepsilon
^C_{\alpha \alpha}$ and $E^{st}_{pn} = \Delta E_{pn} - 3\varepsilon ^C_{pn\alpha}$.
The values are presented in Table 1. This way to obtain empirical values of $E^C$
and $E^{st}$ makes the total binding energy equal to $E^b$ (1)/(2), see Table 1.

The formula to calculate the nuclear binding energy is written as follows
\begin{equation}
E^b = E_{\alpha}^{nuc} - E^C + E^{st}+ E_{\Delta N},\label{7}
\end{equation}
For the lightest nuclei with $Z \leq 8$, $N_\alpha \leq 4$, $E^{st}$=0. The values
of $E_{\alpha}^{nuc}$ and $E^C$ in this case are calculated in accordance with Fig.
1 by adding the corresponding energy portions of the $\alpha$-clusters and the
$pn$-pair. In case of the nuclei not having a core the binding energy of excess
neutrons $E_{\Delta N}$ stays out of consideration.

For the other nuclei with $Z \geq 9$ $E_{\alpha}^{nuc}$ is calculated by (5), $E^C$
is calculated by a well known formula for a spherical charged body with radius
$R_C$, for odd $Z$ nuclei $R_{C1}$ ($R_C$ and $R_{C1}$ are determined by (22) on the
number of $\alpha$-clusters in the nucleus, see Section 3)
\begin{equation}
E^C  = \frac{3}{5}\frac{Z^2e^2}{R_{C}}, \label{8}
\end{equation}
$E^{st}$ is calculated by (6) with $E_\alpha^{st}=\gamma_1 R_{C}^2$, so Eq.(6) is
rewritten as the following
\begin{equation}
E^{st} = E^{st}_5 + \sum^{N_\alpha}_6\gamma_1 R_{C_i}^2; E^{st}_1 = E^{st} +
\gamma_1 R_{p1}^2/2,\label{9}
\end{equation}
where $E^{st}_5 = 1.026$ MeV (see Table 1 for $Z =10$, $N_\alpha=5$), $\gamma_1 =
0.471$ $\rm MeV/fm^2$, $R_{p1}$ stands for the radius of the $pn$-pair (21), see
Section 3. The binding energy of excess neutrons $E_{\Delta N}$ is calculated by
(13), see Subsection 2.2.

The values of $E^b$ (7) for the nuclei with $N=Z$, $E_{\Delta N} =0$, are presented
in Table 1 in the column of '$\Delta E^b$(7)'.

The well known Weizs\"{a}cker formula [20] is
\begin{equation}
E^{Wz} = \alpha A - \beta A^{2/3} - \frac{\gamma Z^2}{A^{1/3}} \pm
\frac{\delta}{A^{3/4}} - \frac{\epsilon(A/2 -Z)^2}{A}.\label{10}
\end{equation}

The values $\Delta E^b$ (7) depends on $N_\alpha$ and $\Delta N$, $\Delta E^{Wz}$
(10) depends on $A$ and $Z$. The least differences between the values of the Coulomb
energy estimated in these approaches are for the nuclei with $N=Z$, see Table 1
[17]. With $Z$ growing the difference increases. Calculation of $E^b$ (7) for
$^{164}_{66}Gd$, $E_{exp}=1338$ MeV,  gives $\frac{3}{5}\frac{Z^2e^2}{R_{C}}= 627$
MeV and the nuclear force energy of all inter-nucleon interactions $E_{\alpha}^{nuc}
+ E^{st}+ E_{\Delta N} =1964$ MeV, so $\Delta E^b = 1$ MeV. Calculation of $E^{Wz}$
(10) gives $\frac{\gamma Z^2}{A^{1/3}}=565$ MeV and $\alpha A - \beta A^{2/3} +
\frac{\delta}{A^{3/4}} - \frac{\epsilon(A/2 -Z)^2}{A}= 1902$ MeV, so $\Delta E^{Wz}
= 1$ MeV. In despite of the fact that $\Delta E^{Wz} = \Delta E^b$, these two
approaches have different values of the Coulomb energy and the nuclear force energy
of all inter-nucleon interactions.

The values of $E^b$ (7) and $E^{Wz}$ (10) calculated for the nuclei with different
number of excess neutrons are presented in Fig 2. Every nucleus is presented in the
graphs '$E^b$' and '$E^{Wz}$' by two dots, one for $E_{exp}$ and the other for the
calculated value.

\begin{figure}[th]
\centerline{\psfig{file=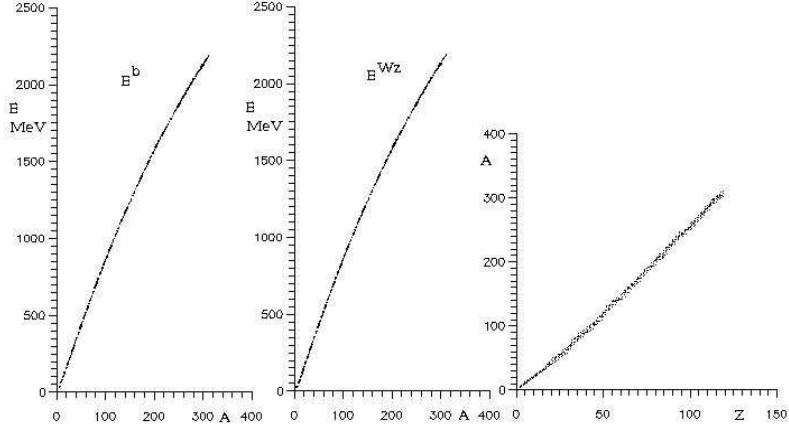,width=14cm}} \vspace*{8pt} \caption{The values of
$E^b$ (7) and the Weizs\"{a}cker formula $E^{Wz}$(10) calculated for the nuclei
$A(Z)$ indicated in the corresponding graph.}
\end{figure}

The widths of the graphs contain both the width of $E_{exp}$ distribution on $A$ and
the deviations between experimental values and calculated ones. For the nuclei with
$Z> 100$ the values calculated by the Weizs\"{a}cker formula have been used instead
of unknown experimental values. The width of graph '$E^b$' in the part corresponding
to the nuclei with $Z > 100$  consists of the both the width of '$E^{Wz}$' and the
deviations $E^{Wz}-E^b$, which is within a few MeV. The width of the part of '$E^b$'
is a little bit bigger than that of the corresponding part of '$E^{Wz}$' graph where
$\Delta E^{Wz} = 0$. The equal widths of the two graphs for the nuclei with $Z \leq
100$ clearly show that the deviations $\Delta E^b$ and $\Delta E^{Wz}$ are close.

To test the validity of formulas (5), (8) and (9) independently of the energy of
excess neutrons, the alpha-particle separation energy $E^{sep}_{\alpha} = E^b(A,Z) -
E^b(A-4,Z-2)$ as well as the deutron separation energy $E^{sep}_{d} = E^b(A,Z) -
E^b(A-2,Z-1)$ for the nuclei with even and odd number of excess neutrons $\Delta N$
has been calculated. In Fig. 3 the calculated values are given in comparison with
the experimental data.

\begin{figure}[th]
\centerline{\psfig{file=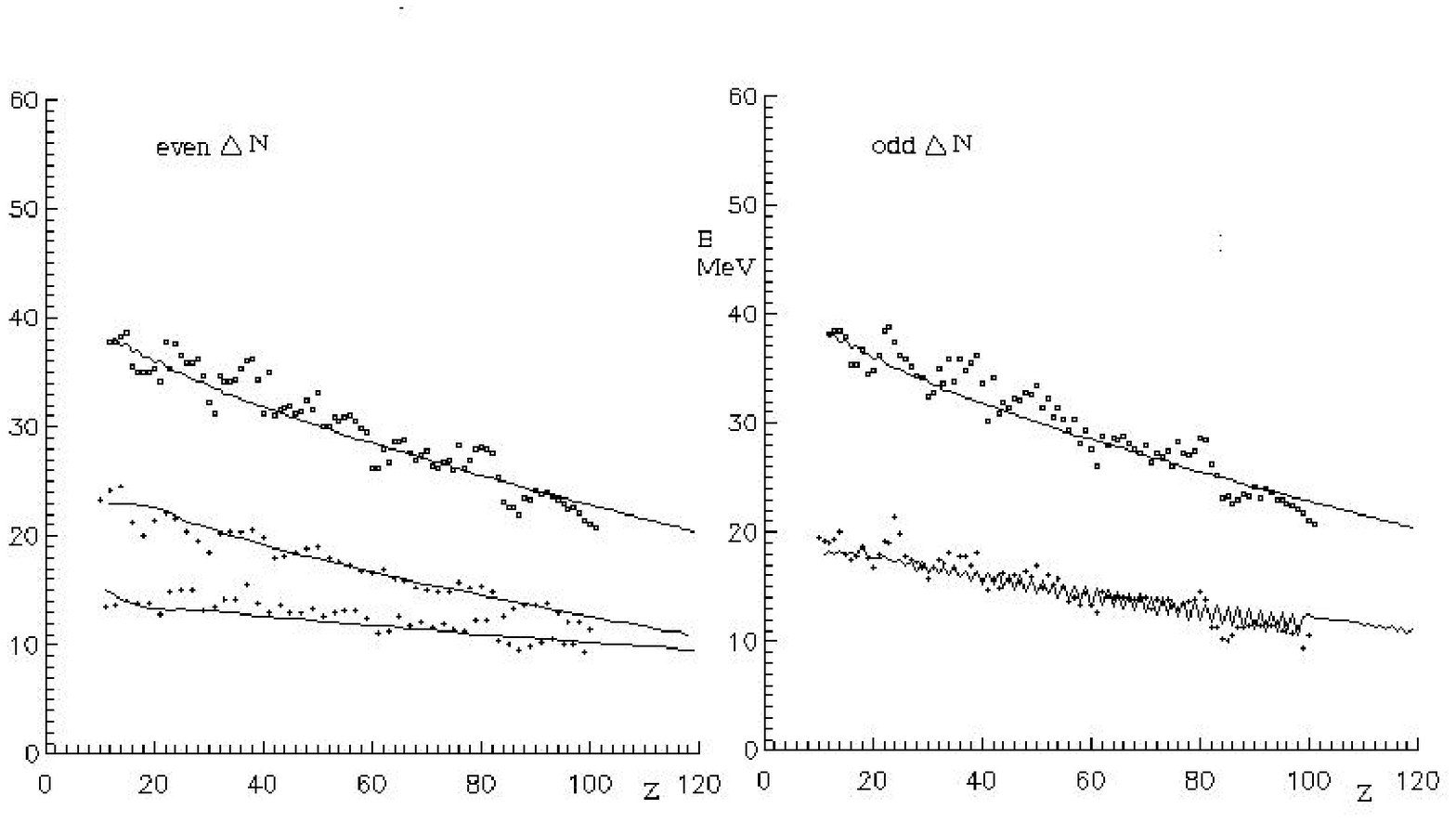,width=12cm}} \vspace*{8pt} \caption{Calculated
values of $\alpha$-particle and deutron separation energy (solid lines) in
comparison with the experimental values of $\alpha$-particle separation energies
(squares) and deutron separation energies (crosses) for the nuclei of
$\beta$-stability path and near it. Left graphs are for the even number of excess
neutrons and the right graphs are for the odd number.}
\end{figure}

The formula to calculate $E^{sep}_{\alpha}$ does not contain $E_{\Delta N}$ neither
for even nor for odd excess neutrons. Neither does the formula to calculate
$E^{sep}_{d}$ in case of even number of excess neutrons.  The deutron separation
energy in the nuclei with even number of excess neutrons splits into two lines in
dependence on whether the nucleus is of even $Z$ or odd $Z$. It is clearly seen in
the left graph of Fig. 3.

In case of odd number of excess neutrons the experimental energy of the single
excess neutron is bigger for odd $Z$ nuclei than that of even $Z$ nuclei. It is
taken into account in the formula to calculate the energy of the last excess neutron
(14), see Subsection 2.2. Then the formula to calculate the deutron separation
energy must have $-0.2E_{nn}$ for even $Z$ and $+0.2E_{nn}$ for odd $Z$ nuclei where
$E_{nn}$ is the binding energy of the $nn$-pair the single neutron belongs to. This
is in agreement with the experimental data, see the right graph in Fig. 3.

One can see a periodical structure in the experimental data, which refers to the
shell effects connected with the spin-orbit correlations of the last nucleons, so
called valent nucleons. The calculated values are not sensitive to the details. They
describe the $\alpha$-particle and deutron separation energies for all nuclei with
an average deviation from the experimental values in 2.5 MeV, which is a small
percentage of the total binding energy.

After simplifying Eqs. (5) and (8) one gets $E_{\alpha}^{nuc}= 42.110
N_\alpha-26.100$ MeV, $E_{\alpha 1}^{nuc}= E_{\alpha}^{nuc} + 16.884$ MeV, and
$E^C=1.848 N_\alpha^{5/3}$, for odd $Z$ $E^C=1.848 (N_\alpha+0.5)^{5/3}$.

For the nuclei with $Z \leq 29$ the empirical values of $E^C$ and $E^{st}$ have been
obtained with a good accuracy, Table 1. It is good to have a simple function to
calculate $E^{st}$ instead of the sum (9). It can be successfully approximated in
accordance with the representation of the nucleus as a core plus four peripheral
clusters like it is used in calculations of radii (24), see Section 3. Then the
function consists of the surface tension energy of the core clusters $E^{st}_{core}$
and  the energy of four peripheral clusters placed at a distance $R_p$ (21) from the
center of mass
\begin{equation}
E^{st} = E^{st}_{core} + 4\gamma_1 R_p^2, E^{st}_1 = E^{st} + \gamma_1 R_{p1}^2/2,
\label{11}
\end{equation}
where $E^{st}_{core} = (N_\alpha- 5.28)(N_\alpha-5)^{2/3} -11.5$ MeV. The
approximation is good for the nuclei with $Z\geq 22$.

A simple function to approximate $E^{st}$ for $Z \geq 30$, $N_\alpha \geq 15$, is
the following
\begin{equation}
E^{st}=(N_\alpha + 1.7)(N_\alpha -4)^{2/3}; E^{st}_1 = (N_\alpha + 2.8)(N_\alpha
-4)^{2/3}.\label{12}
\end{equation}

The  values of $|\Delta E^b|$ of $E^b$ (7) with using $E^{st}$ (9), (11) and (12) as
well as the values of  $|\Delta E^{Wz}|$ for the most of the cases are less than
0.5\% of the total binding energy.

\subsection{Excess neutron binding energy}

As it was shown [10] the core starts growing with $N_\alpha$ from the nucleus with
$N_\alpha =5$. It is suggested that the value of the binding energy $E_{\Delta N}$
of the number of excess neutrons $\Delta N$ placed in the core and bound in $M$
pairs, $\Delta N=2M+1$is calculated as follows:
\begin{equation}
E_{\Delta N} = \sum_{i=1}^{M}E_{nn_i} + \epsilon_n,\label{13}
\end{equation}
where $E_{nn_i}$ stands for the energy of $i$th $nn$-pair, $\epsilon_n$ is the
energy of the single excess neutron and
\begin{equation}
\epsilon_n = E_{nn_{M+1}}/2 \pm 0.1E_{nn_{M+1}},\label{14}
\end{equation}
where '+' stands for odd $Z$ nuclei and '-' for even $Z$ ones. The phenomenological
formula takes into account that fact that a sum of experimental energies of single
excess neutrons  of the nuclei with odd $Z$ and even $Z$ approximately equals the
energy  $E_{nn_{M+1}}$ of the $M+1th$ $nn$-pair which they belong to. In case of
$\Delta N = 1$, $M=0$ $E_{nn_{M+1}} = E_{nn_1}$.

The experimental values of $E_{nn_i}$ in MeV are fitted by an equation
\begin{equation}
E_{nn_i} = 22.5 - 1.358 i^{2/3}. \label{15}
\end{equation}

It is known that the experimental values of separation energy of $nn$-pairs have
deviations within up to 5 MeV determined by  what nucleus loses the pair, which
certainly refers to the shell effects. Despite the relatively big deviation of the
values, the empirical values of the energies of  all excess $nn$-pairs $E_{\Delta
N(exp)}$ known only  for the nuclei with $N = Z$ from the following equation:
\begin{equation}
 E_{\Delta N(exp)}  = E_{exp}(Z, N + \Delta N) - E_{exp}(Z,N), \label{16}
\end{equation}
are restricted within relatively narrow corridor. For example, for the nuclei with
$21 \leq Z \leq 29$ the value of the separation energy of two excess neutrons varies
within $E_{2(exp)} = 21 \div 23$ MeV, for four excess neutrons $E_{4(exp)} = 42 \div
45$ MeV, $E_{6(exp)} = 61 \div \rm 63$ and $E_{8(exp)} = 77 \div 80$ MeV. Therefore,
the parameters  in (15) have been obtained by fitting both the experimental values
of $nn$-pair separation energies known [18] for 27 $nn$-pairs and the values of
$E_{\Delta N(exp)}$. The values of $E_{\Delta N}$ (13) have been used here to
calculate the binding energies of the excess neutrons filling the core, which exists
only in the nuclei with $ Z \geq 10$, $N_\alpha \geq5$. It is evident that Eq. (13)
is not proper for the exotic nuclei like, for example $^6\rm He$ or $^{11}\rm Li$.

\section{Radius of the last proton position, Coulomb radius  and charge radii of nuclei}

To estimate the value of the  radius of the last proton position $R_p$ in a nucleus
$A$ the value of $\Delta E_{pn}$ is taken as the Coulomb energy of the last proton.
The center of mass in the approach is supposed to be in the center of the Coulomb
field. The energy of the last proton in the nucleus with an even $Z$ consists of the
energy of its interaction with the other proton of the $\alpha$-cluster $\varepsilon
^C_\alpha$ and the Coulomb energy of its interaction with the other $Z - 2$ protons
of the nucleus

\begin{equation}
\Delta E_{pn} = \varepsilon ^C_\alpha +(Z-2)e^2/R_p, \label{17}
\end{equation}
and for odd $Z$

\begin{equation}
\Delta E_{pn} = (\varepsilon^C_{pn \alpha} - \varepsilon ^C_{\alpha}) + (Z-1)e^2/
R_{p1}, \label{18}
\end{equation}
where $R_{p1}$ stands for the radius of the position of the single $pn$-pair.

It is valid only for the nuclei with $Z\geq 10$, because  the last proton Coulomb
energy for the lightest nuclei must be calculated with a different function. For
example, one can use the spherical function with the Coulomb radius $R_C$ like the
Coulomb potential at the small radii less than $R_C$ in the Shr\"{o}dinger equation
for proton bound state wave function in a Distorted Wave Born Approximation.

Another way to estimate $R_p$ comes from an equation for the Coulomb energy for the
last $\alpha$-cluster. The Coulomb energy of the cluster consists of the internal
Coulomb energy of the cluster $\varepsilon ^C_\alpha$, Coulomb energy of its
interaction with the three nearby clusters $3\varepsilon ^C_{\alpha \alpha}$ and the
long-range part of the Coulomb energy of its interaction with the other $N_\alpha -
4$ clusters of the nucleus

\begin{equation}
\Delta E_\alpha = \varepsilon ^C_\alpha  + 3\varepsilon ^C_{\alpha \alpha}  +
2(Z-8)e^2/ R_{N\alpha-4}, \label{19}
\end{equation}
where $R_{N\alpha-4}$ stands for the distance between the mass center of the remote
$N_\alpha - 4$ clusters and  the cluster under consideration. The value
$R_{N\alpha-4}$ is approximated as $R_{N\alpha-4} = 1.2R_p$ by fitting the empirical
values of $\Delta E_\alpha$. For the nuclei with odd $Z$ in accordance with the same
logic of taking into account the long-range Coulomb interaction, one obtains

\begin{equation}
\Delta E_{pn} = 3\varepsilon ^C_{pn\alpha}  +  (Z-7)e^2/ (1.2R_p+R_{p1}-R_p).
\label{20}
\end{equation}

The third way to estimate $R_p$ comes from the idea that the long-range Coulomb
energy must be compensated by the surface tension energy $E^{st}_\alpha$. The latter
is expected to be proportional to $R_p^2$. Therefore the last member in the sum (19)
is taken equal to $E^{st}_\alpha = \gamma_1 R_p^2 = 2(Z-8)e^2/ (1.2 R_p)$, and the
value of $\gamma_1 = 0.471\rm MeV/fm^2$ is obtained from fitting the empirical
values of $\Delta E_\alpha$. As a result, one obtains an equation to calculate $R_p$
in dependence on $N_\alpha$ only.

\begin{equation}
R_p=2.168 (N_\alpha-4)^{1/3}.  \label{21}
\end{equation}
The value of $R_{p1}$ is calculated by the same formula  with using $N_\alpha+0.5$
instead of $N_\alpha$.

The values of $R_{p/p1}$ (21) have been used in (17)/(18) to estimate the values of
$\Delta E_{pn}$ for the nuclei with $Z \geq 23, 26, 27, 28$. The values are given in
Table 1. The accuracy of several tenth MeV is determined by the deviation between
two possible ways to obtain $\Delta E_{pn}$ by (17)/(18) or (19)/(20).

Finally, the fourth way to calculate $R_p$ comes from the requirement that the
surface tension energy is expected to be proportional to the number of clusters on
the surface of the liquid drop  $E^{st}_\alpha= \gamma_1 R_p^2 = \gamma_2 N_\alpha
^{2/3}$, where $\gamma_2 =1.645$ MeV is obtained by fitting the values of $\Delta
E_\alpha$. The value is supposed to refer to the Coulomb radius $R_C$. Therefore the
equation  is

\begin{equation}
R_C=1.869 (N_\alpha)^{1/3}.\label{22}
\end{equation}
For odd $Z$ the value of  $R_{C1}$ is calculated by (22) with $N_{\alpha 1}=N_\alpha
+ 0.5$.

The values $R_{p/p1}$  obtained by different ways differ within 0.3 fm for the
nuclei with $16 \leq Z \leq 29$. For the nuclei with $9 \leq Z \leq 15$ the
difference is bigger. For the nuclei with $Z > 22$, $N_\alpha > 11$, $R_C (22) <
R_p$ (21).

It was shown [10] that the formula to calculate the radii of nuclei
\begin{equation}
R = R_\alpha (N_\alpha)^{1/3},\label{23}
\end{equation}
successfully fits the experimental radii of the most abundant isotopes
$R^{abn}_{exp}$ with three close values of $R_\alpha$. Using the value of
$R_{\alpha}$ equal to $R_{\rm ^4 He} = 1.710$ fm [22] fits the radii of the nuclei
with few clusters $N_\alpha = 3, 4$. There is no core in these nuclei. In the nuclei
with $5 \leq N_\alpha \leq 12$, where the number of clusters in the core is
comparable with the number of peripheral clusters,  the value $R_{\alpha} = 1.628$
fm. In the nuclei with $N_\alpha \geq 12$ the radii $R^{abn}_{exp}$ are well
described by the radius  $R_{\alpha} = 1.600$ fm, which is the radius of the
clusters of the core [10].

Varying the number of the peripheral clusters from 1 to 4 (or 5) in accordance with
the comprehensive shell model is the point of the alpha-cluster shell model to
calculate the root mean square radius $R_{shl}$ [14,17]. Then the formula is similar
to that usually used in the framework of the single-particle potential approaches
with using occupation numbers assumed in the shell model [23] in calculations of
root mean square radii.

It was also shown [17] that the number of the peripheral clusters can be fixed at 4
for the nuclei with $N_\alpha \geq 12$, then the root mean square radius $R \approx
R_{shl}$ where $R$ is calculated as follows

\begin{equation}
N_{\alpha } R^2 = (N_{\alpha }-4) 1.600^2 (N_{\alpha }-4)^{2/3} + 4 R_{p}^2.
\label{24}
\end{equation}

For odd $Z$ the charge radius $R_1$ is calculated by the equation

\begin{equation}
(N_{\alpha} + 0.5) R_1^2 = N_{\alpha } R^2 + 0.5 R_{p1}^2, \label{25}
\end{equation}
where the radius of the position of the single $pn$-pair $R_{p1}$ (21) is weighted
by factor 0.5. For the nuclei with $5 \leq N_\alpha \leq 10$ $R$ is calculated as
$R_{shl}$, see (15) in [17].

The radii calculated in the representation of the alpha-cluster shell model
$R_{shl}$ have deviation $<\Delta^2>^{1/2} = 0.034$ fm with (24) (for odd $Z$ (25))
for the nuclei with $9 \leq Z \leq 118$ [17].

A good agreement between the values of R (23) and $R$ (24) (for odd $Z$ (25)) and
the experimental radii of the most abundant isotopes $R^{abn}_{exp}$ [9,22,24] is
shown in Fig. 4. From the figure one can see that for all nuclei with $Z \geq 10$,
$N_\alpha \geq 5$, the Eqs. (23) and (24) give close to the experimental data
values. The values as well as the calculated radii of the $\alpha$-decay nuclei
stable to $\beta$-decay [21] with $83 \leq Z\leq 117$ are presented in the tables in
[10].

\begin{figure}[th]
\centerline{\psfig{file=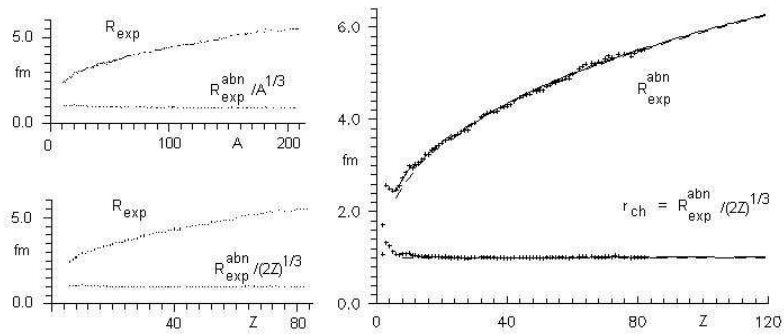,width=11cm}} \vspace*{2pt} \caption{Left part:
Experimental radii of stable isotopes $R_{exp}$  in dependence on $A$ and $Z$. On
the right part: Charge radii $R$ (23) for the $5 \leq Z \leq 23$ and (24)/(25) for
the $Z \geq 24$  (solid line) and $R=1.600 N_\alpha^{1/3}$ fm (dashed line) in
comparison with the experimental radii of the most abundant isotopes $R^{abn}_{exp}$
(crosses). The charge radius of one nucleon of $\alpha$-clusters $r_{ch} =
R/(2Z)^{1/3}$ is indicated by the lines corresponding to $R$.}
\end{figure}

The values of $R$ calculated for two nuclei $^6\rm Li$, $^9\rm Be$ are not in
agreement with the experimental data. These nuclei do not have enough number of
bonds between the clusters and the single $pn$-pair to provide a sufficient nuclear
density. For the nuclei with $6 \leq Z \leq 83$ the deviation of $R$ (23) and
(24)/(25) from the experimental values $R_{exp}$ is  $<\Delta ^2>^{1/2} = 0.050$ fm.
The deviation between the two lines in Fig. 4 for the nuclei with $24 \leq Z \leq
118$ is $<\Delta ^2>^{1/2} = 0.028$ fm. The values of experimental radii of 125
stable isotopes $R_{exp}$ [9,22,24,25,26] are also given  in dependence of $A$ and
$Z$. The radii $R_{exp}$  are given without indication of their errors, which in
most of the cases are within $0.004 \div 0.060$ fm. It visually approves that fact
[10] that both distributions of radii on $A$ and $Z$ are well described by the
functions $\sim A^{1/3}$ and $\sim Z^{1/3}$.

\section{Conclusions}

The approach based on the $\alpha$-cluster model proposes some formulas to calculate
the binding energies and the charge radii of the nuclei of the $\beta$-stability
path and around it. The formulas have been derived on the basis of the idea of
isospin independence of inter-nucleon interactions in the $\alpha$-cluster
structure.

The approach implies that the nucleus is a dense package of alpha-clusters. The
inter-cluster distances are determined by the charge radii of the clusters. Some
amount of excess neutrons fill in the gap between the matter bodies of the
$\alpha$-clusters of the core [10]. The energy of these excess neutrons is described
by a smooth function on the number of the $nn$-pairs (13). The formula to calculate
the binding energy proper for the nucleus with five $\alpha$-clusters (7) turned out
to be good for the other nuclei up to the most heavy ones. The same thing one can
say about the formula to calculate the charge radii $R= r_{ch}(2Z)^{1/3}$ with the
average charge radius of one nucleon of the $\alpha$-cluster structure $r_{ch} \sim
1.01$ fm for the all nuclei with $N_\alpha \geq 5$, see Fig 4.

The formula to calculate the nuclear binding energy is evidently different from the
well known Weizs\"{a}cker formula. These two approaches give different estimations
of the total Coulomb energy and the energy due to all inter-nucleon interactions,
but the values of the total binding energies of these approaches are close. To
calculate the charge radii  both the approaches propose successful but different
formulas, one is $\sim A^{1/3}$ and the other $\sim Z^{1/3}$.

A few useful phenomenological formulas have been found in the approach. These are
the formulas to calculate the root mean square charge radius, the Coulomb radius and
the radius of the last proton's position in dependence on the number of
$\alpha$-clusters. Besides, the empirical values of the Coulomb energy and the
surface tension energy with a good accuracy  have been obtained for the nuclei with
$N=Z$.

The values of the Coulomb repulsion energy and the nuclear force energy of the
inter-cluster interaction, which are 1.925 MeV and 4.350 MeV, correspondingly, could
be a good test in a microscopic description of the light nuclei $^{12}\rm C$ and
$^{16}\rm O$ in the framework of the few-body approach.

\section*{Acknowledgments}

I would like to thank Prof. V. B. Belyaev for helpful discussions the results of the
work.

\end{document}